\documentclass[twocolumn,preprintnumbers,amsmath,amssymb]{revtex4}
\usepackage{graphicx}
\usepackage{epsfig}
\usepackage{amsmath}
\usepackage{amsfonts}
\usepackage{amssymb}
\usepackage{textcomp}
\usepackage{dcolumn}

\begin{document}

\title{About the Origin of the Division between Internal and External Symmetries in Quantum Field Theory}

\author{Martin Kober}
\email{kober@th.physik.uni-frankfurt.de}

\affiliation{Institut f\"ur Theoretische Physik, Johann Wolfgang Goethe-Universit\"at, Max-von-Laue-Str.~1,
60438 Frankfurt am Main, Germany}

\begin{abstract}
It is made the attempt to explain why there exists a division between internal symmetries referring to quantum numbers and external symmetries referring to space-time within the description of relativistic quantum field theories. It is hold the attitude that the symmetries of quantum theory are the origin of both sorts of symmetries in nature. Since all quantum states can be represented as a tensor product of two dimensional quantum objects, called ur objects, which can be interpreted as quantum bits of information, described by spinors reflecting already the symmetry properties of space-time, it seems to be possible to justify such an attitude. According to this, space-time symmetries can be considered as a consequence of a representation of quantum states by quantum bits. Internal symmetries are assumed to refer to relations of such fundamental objects, which are contained within the state of one single particle, with respect to each other. In this sense the existence of space-time symmetries, the existence of internal symmetries and their division could obtain a derivation from quantum theory interpreted as a theory of information.
\end{abstract}

\maketitle

\section{Introduction}

In the framework of relativistic quantum field theories the standard model of particle physics is based on elementary particles are described as irreducible representations of the Poincare group \cite{Wigner:1939, Weinberg:1995mt}. The way the Poincare group is represented depends on the spin of the particle. However, since the particles are further specified through further quantum numbers, which are not related to space-time there arises the existence of what are called internal symmetries.
All existing interaction theories including gravity can be formulated as gauge theories with respect to local symmetries. This way of formulation provides a very important connection between general relativity and the interactions of the standard model. However, there arises the difference that the symmetry groups of the standard model refer to quantum numbers \cite{Yang:1954,Weinberg:1996kr} and as such represent as already mentioned internal symmetries and the gauge symmetry of general relativity as gauge theory of local translations or Lorentz transformations respectively refers to space-time \cite{Hehl:1976kj,Cho:1975dh}. There can be asked the question if it is possible to infer the existence of space-time symmetries and the existence of internal symmetries from a unified principle. If this would be possible, one would get nearer to the aim of unifying general relativity with quantum field theory and the standard model. Within the approach of Carl Friedrich von Weizsaecker's reconstruction of physics \cite{Weizsaecker:1971, Weizsaecker:1985, Weizsaecker:1992, Lyre:1994eg, Lyre:1995gm, Lyre:2003tr} the existence of a (3+1)-dimensional space-time manifold is the consequence of the representation of physical objects described by quantum states in an abstract Hilbert space $\mathcal{H}^m$ as a tensor product of objects within two dimensional Hilbert spaces $\mathbb{C}^2$, which he calls ur objects (the denotation ur object is derived from the German prefix ur-, which means something like original, elementary or primordial)

\begin{equation}
\mathcal{H}^m \subseteq T^n=\bigotimes_n \mathbb{C}^2,\quad m<2^n.
\end{equation}
This means nothing else than the fact that the information, which is contained in a quantum state is resolved into quantum bits. The fundamental objects, which are represented by single quantum bits, do not presuppose a position space. This is in accordance with the basic postulates of quantum theory in the general Hilbert space formulation of Dirac and von Neumann describing states represented by vectors in a Hilbert space without mandatory referring to position space or to space-time \cite{Dirac, Neumann}.
Within this abstract setting, quantum theory does not yet refer to special physical concepts but it represents a certain kind of logical structure and as such refers to information.  
However, elements of two dimensional Hilbert spaces, which are described by Weyl spinors (being the most fundamental states which are even thinkable in any quantum theoretical description !), already reflect the symmetry properties of real space-time, as a (3+1)-dimensional Minkowski space-time. This correspondence between a two dimensional space of Weyl spinors and the Minkowski space-time has its origin in the isomorphism between the Lorentz group $SO(3,1)$ and the group of general linear transformations in a two dimensional complex vector space $SL(2,\mathbb{C})$. The twistor approach to general relativity of Roger Penrose is based on this isomorphism \cite{Penrose:1985jw, Penrose:1986ca, Penrose:1960eq, Penrose:1977in}. Besides, the subgroup of unitary transformations conserving probability distributions in two dimensions $SU(2)$ corresponds to the group of rotations in three dimensional space $SO(3)$. The ur objects, which represent nothing else than quantum bits, are assumed to be the basic constituents of any physical object or physical system. Under this assumption it does not change the physics if every ur object a physical system consists of is transformed with the same element of the $SL(2,\mathbb{C})$ or the $SU(2)$ symmetry group respectively. Therefore the fact that physical objects can be described and perceived within a uniform position space with rotation and Lorentz symmetry could indeed be assumed to have its origin in the possibility to represent every quantum state as a tensor product of ur objects.
Thus it seems to be possible to hold the attitude that physics can be described by particles represented by states in a position space obeying certain external symmetries because these symmetries already reflect the symmetries of an abstract state in a Hilbert space representing quantum information, which is assumed to be the fundamental entity of nature. This means that the existence of a 3-dimensional position space could be derived from quantum theory. In this sense information would be assumed to be more fundamental than matter and space. The whole attitude obtains strong support by the fact that it is possible to obtain free quantum theoretical field equations in Minkowski space within this approach. 
According to von Weizsaecker's approach, beginning from one single ur object, it is possible to obtain field equations of quantum fields in Minkowski space-time by only presupposing the basic laws of general quantum theory. In \cite{Weizsaecker:1985}
there is obtained the Weyl equation. It shall first be given a short review of this derivation, but it is the aim of this paper to make the attempt to incorporate internal symmetries, or at least the isospin symmetry, to the quantum theory of ur objects. This is done by derivation of the Dirac equation with a mass term.

\section{Derivation of free field equations}

An ur object or a quantum bit described by a two dimensional spinor is obtained by quantising a binary alternative
$a=(1,2)$, which is performed by assigning complex values to the two possible values of the alternative. If the spinor 
representing the ur object is denoted as u, it can be mapped to a Minkowski vector $k^\mu$ in the following way

\begin{equation}
k^\mu \equiv u^{\dagger} \sigma^\mu u,
\label{spinor_Minkowskivector}
\end{equation}
where the $\sigma^\mu$ describe the Pauli matrices the unity matrix in two dimensions included.
This vector is a lightlike vector and it obeys the following algebraic relation 

\begin{equation}
k_\mu \sigma^\mu u=0.
\label{Weyl_constraint}
\end{equation}
The Minkowski vector $k^\mu$ can now be seen as a classical quantity with respect to a further quantization procedure. Such a quantization corresponds to an assignment of complex values to the possible values of the Minkowski vector leading to a wave function depending on the Minkowski vector $\varphi(k^\mu)$

\begin{equation}
k^\mu \rightarrow \varphi(k^\mu).
\label{quantization_minkowski_vector}
\end{equation}
It is now possible to define a spinor wave function in the following way

\begin{equation}
\psi(k^\mu)\equiv u \varphi(k^\mu).
\end{equation}
Because of the relation ($\ref{Weyl_constraint}$), only such Minkowski vectors are allowed to have values of the wave function unequal to zero, which obey the constraint

\begin{equation}
k_\mu \sigma^\mu \psi(k^\mu)=0,
\end{equation}
which is implemented in the sense of Dirac. Performing a Fourier transformation

\begin{equation}
\psi(x^\mu)=\int d^4 k e^{i k_\mu x^\mu} \varphi(k^\mu)
\end{equation}
by introducing the coordinate $x^\mu$ leads to the Weyl equation

\begin{equation}
i \sigma^\mu \partial_\mu \Psi(x^\mu)=0,
\label{Weyl_equation}
\end{equation}
if the vector $k_\mu$ is identified with a four momentum and the parameter $x_\mu$ is identified with a position in real space-time. That the vector $x_\mu$ can be identified with a point in real space-time is in accordance with the translation gauge invariance of ($\ref{Weyl_equation}$) with respect to $x^\mu$. Since a momentum $k^\mu$ corresponds according to ($\ref{spinor_Minkowskivector}$) to a single ur object, a state with a sharp momentum, call it $p^\mu$, $\varphi(k^\mu)=\delta(k^\mu-p^\mu)$, which is completely delocalized in position space, contains only one quantum bit. To obtain a state, which is localized very sharply in position space, one has to superpose many wave functions with sharp momentum. According to this, a state contains the more information the more sharply it is localized in position space. 

\section{Derivation of the Dirac equation and introduction of mass}

So far it has been derived the Weyl equation, a free quantum theoretical field equation of a massless particle. There arises the question how masses of particles can be introduced. But even more important seems to be the question about the internal symmetries related to quantum numbers like the isospin for example. 
If the symmetry property of the Hilbert space of quantum objects represented by ur objects is already related to the space-time symmetries, there arises the question how the existence of the internal symmetries can be explained, which have to be related to the symmetry properties of the ur objects within this approach, too. 

It is possible to obtain a field equation with a mass term, the Dirac equation namely, if the Minkowski vector being identified with a momentum in Minkowski space-time is constructed from two ur objects, call them u and v, which can in accordance with the two different representations of the Lorentz group within a spinor space be incorporated to a single Dirac spinor

\begin{equation}
\chi \equiv \left(\begin{matrix} u \\ i \sigma^2 v^{*}\end{matrix}\right) \equiv \left(\begin{matrix} \chi_A \\ \chi_B \end{matrix}\right).
\end{equation}
This Dirac spinor can be mapped to a Minkowski vector according to

\begin{equation}
k^\mu \equiv \bar \chi \gamma^\mu \chi=\chi_A^{\dagger} \sigma^\mu \chi_A-\chi_B^{\dagger} \sigma^\mu \chi_B,
\label{Diracspinor_Minkowskivector}
\end{equation}
where the $\gamma^\mu$ describe the Dirac matrices and $\bar \psi$ is defined as $\psi^{\dagger} \gamma^0$.
Such a Minkowski vector constructed from a Dirac spinor is not mandatory lightlike. It fulfils the relation

\begin{equation} 
k_\mu k^\mu=m^2,
\label{relation_mass}
\end{equation}
with 

\begin{eqnarray}
m& \equiv &2\left(\chi_{A1}^{*}\chi_{A1}\chi_{B2}^{*}\chi_{B2}+\chi_{A2}^{*}\chi_{A2}\chi_{B1}^{*}\chi_{B1}\right. \nonumber\\
&&\left. -\chi_{A1}^{*}\chi_{A2}\chi_{B2}^{*}\chi_{B1}-\chi_{A2}^{*}\chi_{A1}\chi_{B1}^{*}\chi_{B2}\right)^{\frac{1}{2}} \nonumber\\
&=&2\left(u_{1}^{*}u_{1}v_{1}^{*}v_{1}+u_{2}^{*}u_{2}v_{2}^{*}v_{2}\right. \nonumber\\
&&\left. +u_{1}^{*}u_{2}v_{2}^{*}v_{1}+u_{2}^{*}u_{1}v_{1}^{*}v_{2}\right)^{\frac{1}{2}}.
\end{eqnarray}
m is equal to zero if the spinors $\chi_A$ and $\chi_B$ show to the same direction and it is unequal to zero if they 
show to opposite directions. This means that if the spinors of the ur objects $u$ and $v$ take the same value

\begin{equation}
u=\left(\begin{matrix}1 \\ 0 \end{matrix}\right) \quad,\quad v=\left(\begin{matrix}1 \\0 \end{matrix}\right)\quad or \quad
u=\left(\begin{matrix}0 \\ 1 \end{matrix}\right) \quad,\quad v=\left(\begin{matrix}0 \\1 \end{matrix}\right), 
\end{equation}
there arises a mass term. But if they take different values

\begin{equation}
u=\left(\begin{matrix}1 \\ 0 \end{matrix}\right) \quad,\quad v=\left(\begin{matrix}0 \\1 \end{matrix}\right)\quad or \quad
u=\left(\begin{matrix}0 \\ 1 \end{matrix}\right) \quad,\quad v=\left(\begin{matrix}1 \\0 \end{matrix}\right), 
\end{equation}
then the right hand site of equation ($\ref{relation_mass}$) vanishes. Linearizing of ($\ref{relation_mass}$) in the usual sense of Dirac leads to

\begin{equation}
\left(\gamma^\mu k_\mu+m \right)\left(\gamma^\nu k_\nu-m \right)=0.
\label{linearization}
\end{equation}
Performing another quantization of the obtained Minkowski vector leads again to a wave function $\varphi(k^\mu)$

\begin{equation}
k^\mu \rightarrow \varphi(k^\mu).
\label{wavefunction_Dirac}
\end{equation}
This time a corresponding spinor wave function has to be constructed by introduction of two further ur objects, call them w and x, from which there can be obtained a further Dirac spinor

\begin{equation}
\psi \equiv \left(\begin{matrix} w \\ i \sigma^2 x^{*} \end{matrix}\right) \equiv \left(\begin{matrix} \psi_A \\ \psi_B \end{matrix}\right).
\end{equation}
A corresponding spinor wave function can be defined according to

\begin{equation}
\Psi(k^\mu)=\psi \varphi(k^\mu).
\label{Dirac_spinor}
\end{equation}
Application of the first factor of the linearization ($\ref{linearization}$) to the spinor ($\ref{Dirac_spinor}$) leads to

\begin{equation}
\left(\gamma^\mu k_\mu+m \right)\Psi(k^\mu)=0
\end{equation}
and performing a Fourrier transformation 

\begin{equation}
\Psi(x^\mu)=\int d^4 k e^{i k_\mu x^\mu} \Psi(k^\mu)
\end{equation}
leads to the celebrated Dirac equation

\begin{equation}
\left(i\gamma^\mu \partial_\mu-m \right)\Psi(x^\mu)=0.
\label{Dirac_equation}
\end{equation}
A further quantization of the state $\Psi(x^\mu)$ 

\begin{equation}
\{\bar \Psi(x,t),\Psi(x',t)\}_+=i\delta(x-x'),
\end{equation}
leads to the usual setting of quantum field theory.
To obtain a state, which is not completely delocalized in position space one needs again many ur objects. However, a state with a sharp momentum consists of four ur objects now. The objects u and v constitute the momentum and the objects w and x determine the direction of the spin and if the state describes a particle or an anti particle. But this is not the complete truth. The interesting thing is the mentioned fact that there are possible massive and massless states and this depends on the relative orientation of the states representing the ur objects u and v, from which the momentum vector is constructed. This enables the possibility to incorporate the isospin.

\section{Incorporation of the Isospin and the division between internal and external symmetries}

Within the standard model the isospin is a quantum number distinguishing states of particles with different mass and different charge \cite{Weinberg:1967tq,Weinberg:1996kr}. So far we are just dealing with free field equations and therefore charge plays no role yet. In the above construction of the Dirac field and its dynamics from ur objects, there can appear a mass term but the mass term can also vanish. This depends on the relation of the states of the ur objects with respect to each other building the Dirac spinor $\chi$, from which a Minkowski vector is constructed according to ($\ref{Diracspinor_Minkowskivector}$), which is interpreted as a momentum vector of a particle in Minkowski space.
Thus it is opened the possibility that the isospin can be interpreted as the orientation of the two ur objects the momentum consists of with respect to each other. This means that the information the two ur objects contain can be represented as a combination of the relative orientation of one ur object with respect to the second one and the absolute orientation of the second ur object. In this sense the relative orientation of the ur objects interpreted as isospin decides if the momentum vector is lightlike or if it is timelike and thus if the corresponding particle is massive. The orientation of these ur objects with respect to other quantum states describes the momentum in the sense of the usual degree of freedom of a particle.
This implies that the isospin is already implicitly contained in the above description of the Dirac equation by ur objects.
However, in contrast to the space-time symmetries it does not appear explicitly in ($\ref{Dirac_equation}$), which is derived from ($\ref{Diracspinor_Minkowskivector}$), ($\ref{wavefunction_Dirac}$) and ($\ref{Dirac_spinor}$). Therefore it shall be given a reformulation of ($\ref{Dirac_equation}$) to make the appearance of the isospin more explicit.
If a different orientation of the two ur objects corresponding to the case $m=0$ is identified with isospin $1/2$ and an opposite orientation of the ur objects corresponding to the case $m \neq 0$ is identified with isospin $-1/2$, this corresponds to the appearance of the isospin within the lepton sector of the standard model \cite{Weinberg:1967tq,Weinberg:1996kr}. Therefore it shall be defined an isospin vector as follows

\begin{equation}
I \equiv \left(\begin{matrix}1-|\langle u|v\rangle|\\ |\langle u|v\rangle|\end{matrix}\right).
\end{equation}
By referring to this isospin vector there can be defined an extended state

\begin{equation}
\Phi(k^\mu) \equiv I \otimes \Psi(k^\mu)=I \otimes \psi \varphi(k^\mu),
\end{equation}
where $\Psi(k^\mu)$, $\psi$ and $\varphi(k^\mu)$ are the quantities defined in ($\ref{Diracspinor_Minkowskivector}$), ($\ref{wavefunction_Dirac}$) and ($\ref{Dirac_spinor}$).
This spinor wave function represents a dublett with respect to the isospin now and obeys the following field equation

\begin{eqnarray}
&&\left(i\gamma^\mu \partial_\mu-\left(\begin{matrix}0 & 0 \\ 0 & m \end{matrix}\right)\right)
\left(\begin{matrix}1-|\langle u|v\rangle|\\ |\langle u|v\rangle|\end{matrix}\right)\otimes \Psi(x^\mu)\nonumber\\
&&=\left(i\gamma^\mu \partial_\mu-m_I \right)\Phi(x^\mu)=0,
\label{Dirac_Isospin}
\end{eqnarray}
whereas the mass matrix $m_I \equiv \left(\begin{matrix}0 & 0 \\ 0 & m \end{matrix}\right)$ of course refers to the isospin space. This is just a reformulation of ($\ref{Dirac_equation}$), which contains the isospin implicitly as already mentioned.  
If the isospin state $+1/2$ is identified with the electron neutrino and the isospin $-1/2$ is identified with the electron
as a massive particle for example, then equation ($\ref{Dirac_Isospin}$) can be written as

\begin{equation}
\left(i\gamma^\mu \partial_\mu-\left(\begin{matrix}0 & 0 \\ 0 & m_e \end{matrix}\right)\right)\Phi = \left(i\gamma^\mu \partial_\mu-\left(\begin{matrix}0 & 0 \\ 0 & m_e \end{matrix}\right)\right)\left(\begin{matrix}\nu_e \\ e \end{matrix}\right)=0,
\end{equation}
with $\Phi \equiv \left(\begin{matrix}\nu_e\\e \end{matrix}\right)$.
Since the neutrino state as well as the electron state contains a full Dirac spinor described by $\psi$ as spin state, there can appear right handed neutrinos as well as left handed neutrinos. Therefore it is not explained in this model why there exist only left handed neutrinos. But if they are assumed to be massless, as it is done here according to the standard model, they can be interpreted as Majorana particles, which means that the right handed particle corresponds the antiparticle of the left handed particle. 
According to the approach advocated in this paper and the above description of the isospin internal degrees of freedom and the corresponding symmetries have their origin in a relation of two or more ur objects contained within one particle with respect to each other. In contrast to this degrees of freedom referring to space-time and the corresponding symmetries arise from the relation of all the ur objects contained in one state of a particle with respect to other states of particles or even, if one wants to express it this way, to the rest of the universe. 
Since the internal degrees of freedom are assumed to represent the relation of several ur objects within one particle state with respect to each other, a transformation belonging to an internal symmetry corresponds to a transformation of some ur objects within the state of a particle with a certain element of the $SU(2)$. In contrast to this a transformation of all ur objects contained in a particle with a certain element of the $SU(2)$ or the $SL(2,\mathbb{C})$ respectively corresponds
to a rotation in position space or a Lorentz transformation respectively.

\section{Program for an incorporation of higher symmetry groups}

So far it has just been obtained the description of the isospin according to its appearance in the lepton sector of the standard model. There arises the question how higher symmetries like the flavour symmetry or the colour symmetry of the quark sector could be included to this approach.
It seems to be very plausible that it is possible to incorporate higher symmetry groups, especially the symmetry groups
$SU(3)_{flavour}$ and $SU(3)_{colour}$ by constructing already the momentum vector with more ur objects. In this sense higher $SU(N)$ symmetries could arise from a combination of several ur objects, from which each obeys a $SU(2)$ symmetry. The appearance of multipletts from dubletts could appear in the usual way through splitting of the composed Hilbert space of several ur objects into symmetric and antisymmetric states like it is represented by the young tableaus.
This is in accordance with the fact that the isospin seems to be more fundamental than other quantum numbers because all particles, the particles of the lepton sector as well as the particles of the quark sector, are dubletts with respect to the isospin, whereas only the quarks are multipletts with respect to the colour degree of freedom for example. 
Already Heisenberg's unified spinor field theory \cite{Heisenberg:1957,Heisenberg:1967,Heisenberg:1974du} was based on the assumption that the $SU(2)$ symmetry of the isospin represents a fundamental symmetry whereas the higher symmetry groups are only approximate symmetries arising as consequences from the more fundamental $SU(2)$ symmetry.
In \cite{Durr:1979fi} there has been made the attempt to build a composite model of leptons and quarks which is based merely on the $SU(2)$ symmetry of the isospin. The division between the quark and the lepton sector for example could have its origin in a combination of two ur objects leading to a triplett and a sigulett. The singulett could represent the lepton sector and the triplett the quark sector with $SU(3)_{colour}$ symmetry, ${\bf 2} \otimes {\bf 2}={\bf 3} \oplus {\bf 1}$. In this sense multipletts with the corresponding symmetry properties would appear as a consequence of combinations of underlying dubletts with $SU(2)$ symmetry.
In the suggested model the isospin symmetry is very closely related to the space-time symmetries and arises from the very beginning. Since the isospin seems to be more fundamental than the symmetries of the strong interactions, 
it is not astonishing that it is more directly related to the space-time symmetries, as it is the case because it describes the relation of the ur objects the momentum vector of a particle consists of.
\\
\\

\section{Summary and Discussion}

It has been suggested that internal symmetries referring to quantum numbers and external symmetries referring to space-time could have the same origin lying in quantum theory. This suggestion arises from the fact that every quantum state, no matter to which special physical object or context it refers, can be represented as a tensor product of quantum bits, which are assumed to represent the fundamental objects of nature, called ur objects. The symmetry properties of the corresponding Hilbert space, a space of two dimensional spinors, reflects already the symmetry properties of space-time. But if the properties of space-time are assumed to be a kind of representation of the symmetry properties of a quantum state in an arbitrary Hilbert space, then there remains the question why there exist internal symmetries, too. According to the attitude of this paper internal symmetries represent properties, which arise from relations of ur objects belonging to one single particle whereas external symmetries correspond to their relation to other particles or even to the rest of the universe. In accordance with this it has been made the attempt to introduce the isospin by deriving the Dirac equation. This derivation was based on the construction of a momentum vector from two ur objects whereas the relation of the orientation of the two ur objects with respect to each other,
whereof it depends if there appears a mass term, is interpreted as the isospin.\\ 
$Acknowledgement$:\\ I would like to thank the Messer Stiftung for financial support.

\end{document}